\documentclass[3p,times,procedia]{elsarticle}
\usepackage{nupha_ecrc}
\usepackage{wrapfig}
\volume{00}
\firstpage{1}
\journalname{Nuclear Physics A}
\runauth{}
\jid{nupha}  

\jnltitlelogo{Nuclear Physics A}

\usepackage{amssymb}

\usepackage[figuresright]{rotating}
\usepackage{amsfonts, amssymb, amsmath, graphicx, comment, bm, slashed, caption, subcaption, dcolumn,color}

\newcommand{\beq}{\begin{eqnarray}}
\newcommand{\eeq}{\end{eqnarray}}

\usepackage[colorlinks=true,linktocpage=true,linkcolor=blue,citecolor=blue]{hyperref}

\begin{document}

\begin{frontmatter}

\dochead{}

\title{Structure of virtual photon polarization in ultrarelativistic heavy-ion collisions}


\author{Gordon Baym,$^{a,b}$ Tetsuo Hatsuda,$^{b,c}$ and Michael Strickland$^d$}
\address{\mbox{$^a$Department of Physics, University of Illinois, 1110
  W. Green Street, Urbana, IL 61801-3080, United States} \\
\mbox{$^b$iTHES Research Group and iTHEMS Program, RIKEN, Wako, Saitama 351-0198, Japan} \\
\mbox{$^c$Nishina Center, RIKEN, Wako, Saitama 351-0198, Japan} \\
\mbox{$^d$Department of Physics, Kent State University, Kent, OH 44242, United States}
}

\begin{abstract}

   Anisotropy in heavy-ion collisions leads to polarization of both direct and virtual  photons,  the latter detected via internal conversion to  dileptons pairs.   Thus measurement of photon polarization probes the anisotropy at all stages of collisions.  In order to characterize the polarization of virtual photons, we derive here the general structure of the photon polarization operator, $\rho^{\mu\nu}$,
in an anisotropic medium, in terms of the four spectral functions, two transverse and one longitudinal, as usual, plus a new spectral function
that reflects anisotropy in asymmetric collisions; and derive the local production rate of dilepton pairs in terms of these four functions.

\end{abstract}


\end{frontmatter}


\section{Introduction}
\label{sec:1}

   Direct photons, both real and virtual, are valuable probes of the dynamics of ultrarelativistic heavy-ion collisions, comparable to the way the cosmic microwave background (CMB) probes the early universe.    In the early universe, momentum anisotropy of photons at their
last scattering with charged particles -- at the time of decoupling of radiation and matter  -- leads to polarization of the CMB, as a consequence of transverse photon polarization being preserved in Thomson scattering \cite{mrees,Samtleben:2007zz}.
Similarly, direct and virtual photons in heavy-ion collisions provide information on initial as well as later stages of the collisions, e.g., deviations from thermal equilibrium in the hot plasma as reflected in the unpolarized photon spectrum  \cite{Schenke:2006yp}.   
Measurement of photon polarization in collisions, most promisingly that of virtual photons, measured in the angular distribution of the dileptons they produce \cite{Hoyer,Bratkovskaya,Shuryak}, provides information on the anisotropy of the quark-gluon plasma.   In this note we summarize the general framework, based on the photon spectral functions in the plasma, for analyzing the angular distribution and thus the polarization of dileptons in terms of the plasma anisotropies; a more detailed analysis is given in Ref.~\cite{virtual}.   

\begin{wrapfigure}[9]{r}{9cm}
\hspace{-8pt}
\includegraphics[width=9.0cm]{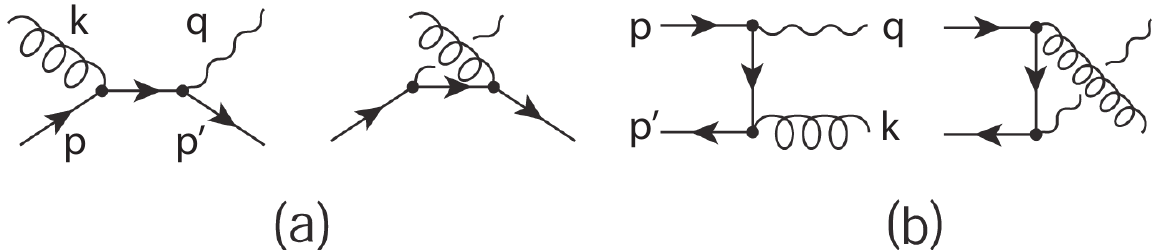}
\caption{
(a) Compton scattering of a gluon against a quark or anti-quark producing a photon; 
(b) production of a photon and gluon by quark-antiquark annihilation. 
}
\label{fig:Compton}
\end{wrapfigure}

     Recently two of us \cite{BH} pointed out how anisotropy in heavy-ion collisions leads to polarization of the direct photons produced in collisions; the principle photon sources are Compton scattering of gluons on quarks and quark-anti quark annihilation into a gluon-photon pair, Fig.~\ref{fig:Compton}.   In general, the photon is polarized along the component of the charge current induced in the collision that is
perpendicular to the photon momentum.  In  Compton scattering of a gluon against a quark at rest, the photon is preferentially  
polarized in the direction perpendicular to the scattering plane, since the charge current is along the direction of the gluon polarization.
Similarly, in pair annihilation of an antiquark with a quark at rest the photon is basically
polarized in the direction parallel to the scattering plane, since the 
charged current is along the incident anti-quark momentum. The two processes tend to produce orthogonal photon polarizations; to see which is dominant requires detailed calculations.   Reference \cite{BH} -- in a schematic model in which the quarks were replaced by heavy scattering centers, with the gluon anisotropy described by an anisotropic temperature and the collision dynamics described by Bjorken expansion -- estimated a net photon polarization of order 10\% or larger.   However,  more realistic calculations \cite{strickland-unpubl} showed the final polarization to be only on the order of 3-5\%.
     
    Furthermore, detection of polarization of direct photons produced in heavy-ion collisions is very difficult.  Essentially one must convert the direct photons in a foil to produce dilepton pairs.   The small opening angle of the pairs plus their subsequent rescattering in the foil makes it
not possible practically  to identify pairs and reconstruct the polarization.  Much more promising is to take advantage of internal conversion of photons into dileptons, whose angular distribution is more readily measured.   The question is how is the anisotropy reflected in the angular distribution of dileptons from internal conversion.     The answer is given in the structure of the photon polarization or self-energy operator to which we now turn.
    
\section{Dilepton production rate and the photon polarization operator}    
    
\begin{wrapfigure}[8]{r}{4cm}
\vspace{-0.5cm}
\hspace{-8pt}
\includegraphics[width=4.0cm]{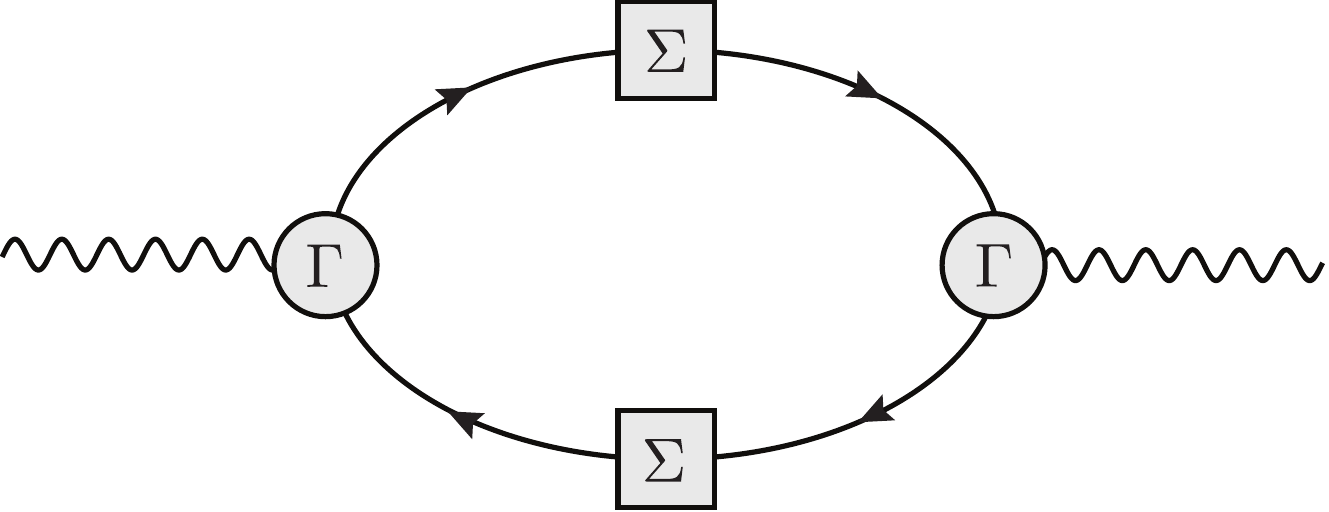}
\caption{
Photon polarization tensor. The $\Gamma$ are hard thermal loop vertex corrections, and $\Sigma$ hard thermal loop quark self-energies.}
\label{bubbles}
\end{wrapfigure}  
    The production rate $R$ of dilepton pairs of 4-momentum $p$ and $p'$ by a virtual photon of 4-momentum $q$, at a given point in space-time in the collision volume, is 
$dR_{l^+  l^-}/d^3\bar{p} d^3\bar{p}'   = (\alpha^2/4\pi^4 Q^4) \  \rho_{\mu\nu} (q)  L^{\mu\nu}(p,p')$,  
where $\rho_{\mu\nu}$ is the spectral function of the in-medium photon self-energy,  illustrated
in Fig~\ref{bubbles}. 
\beq
\Pi_{\mu\nu}(\vec q\,,z) = e^2 \int_{-\infty}^{\infty}\frac{dq^0}{2\pi} \frac{\rho_{\mu\nu}(\vec q\,,q^0)}{z-q^0} \, ,
\eeq
and $L^{\mu\nu} (q,s) =2  \left( q^\mu q^\nu -g^{\mu\nu} Q^2 -s^\mu s^\nu \right)$ is the squared matrix element for a virtual photon of 4-momentum $q$ to produce the lepton pair.  Here $d^3\bar{p} \equiv  d^3p/2E_p$, $d^3\bar{p}' \equiv  d^3p^\prime /2E_{p'}$, $q=p+p'$,
$s=p-p'$, $Q^2 \equiv q^\mu q_\mu > 0$, $g^{00}=1$, $Q^2+s^2=4m ^2$, and $m$ is the lepton mass.

\begin{wrapfigure}[13]{r}{6cm}
\vspace{-1cm}
\hspace{-2cm}
\includegraphics[width=7.5cm]{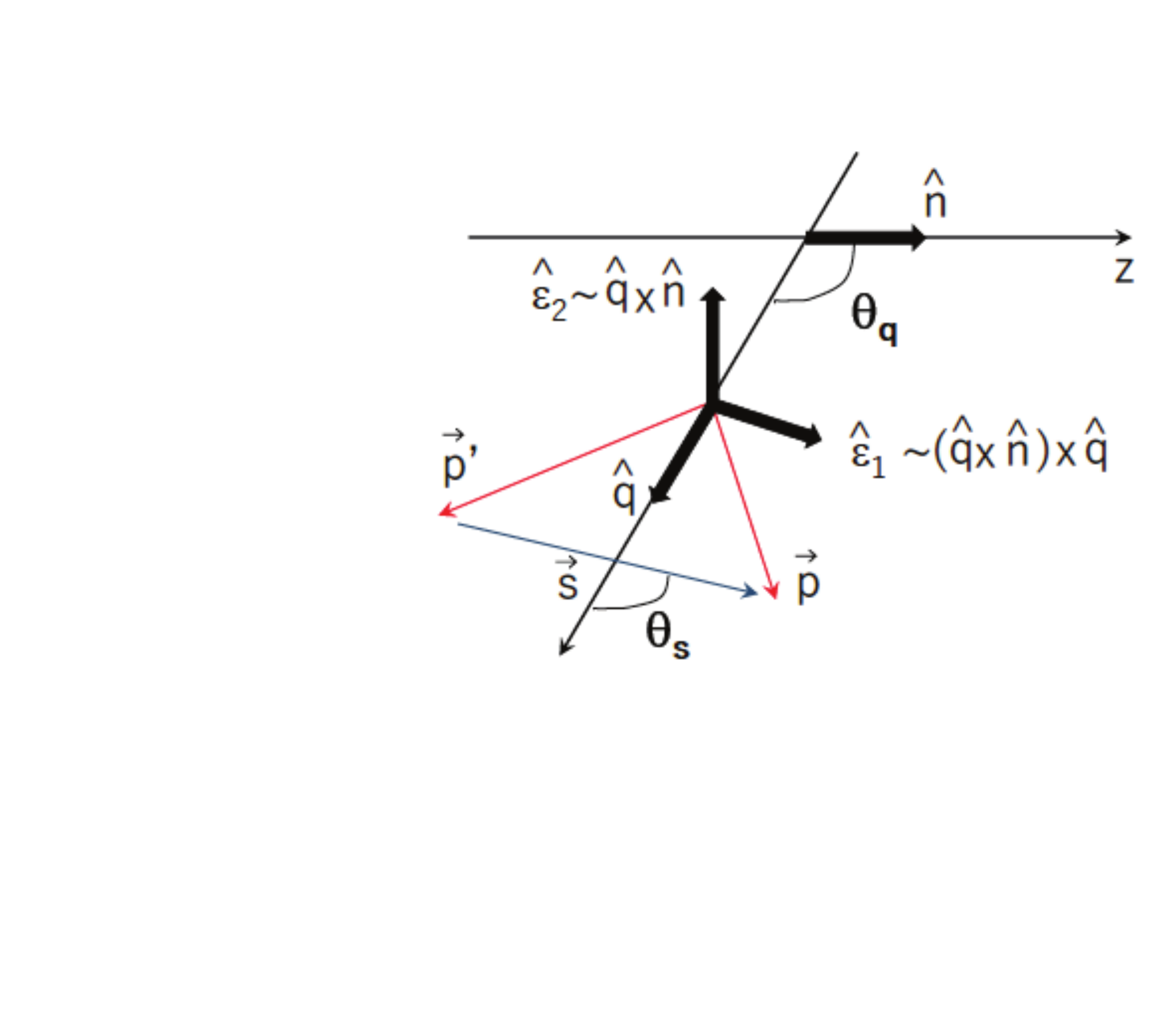}
\vspace{-2cm}
\caption{
Collision geometry showing the polarization vectors $\varepsilon_1$ and $\varepsilon_2$, the anisotropy axis $\hat n$, virtual photon momentum $\vec q$, and the dilepton momenta $\vec p$ and $\vec p'$.}
\label{angles}
\end{wrapfigure} 

  To determine the structure of $\rho_{\mu\nu}(q)$ in an anisotropic medium,  we first construct a basis of polarization vectors, working in the 
local rest frame of the matter, and assuming that the anisotropy is along one axis $\hat n$, generally the beam axis.  We define the two polarizations transverse to the photon momentum $q$:
$\varepsilon_i^\mu = (0, \hat \varepsilon_i)$,
where 
$\hat\varepsilon_1 \equiv (\hat q \times \hat n) \times \hat q/|\hat q \times \hat n|$ and 
 $\hat \varepsilon_2 \equiv \hat q \times \hat n/|\hat q\times \hat n|$.    In addition, we define the longitudinal polarization vector 
$
   \varepsilon_{\rm L}^\mu \equiv  (|\vec {\tilde q}\,|,\tilde q^0\hat q) \, .
$
where $\tilde q^\mu \equiv q^\mu/\sqrt{Q^2}$.  These three polarization vectors have norm $\varepsilon_i^\mu\varepsilon_{i\mu} = -1$, as does the 4-vector axis $n^\mu \equiv (0,\hat n)$.  The geometry of the polarization vectors and dilepton momenta are shown in Fig.~\ref{angles}. 
 Together with $q^\mu$ the polarization vectors form an orthonormal basis obeying
$ g^{\mu\nu} = \tilde q^\mu \tilde q^\nu - \varepsilon_1^\mu \varepsilon_1^\nu - \varepsilon_2^\mu\varepsilon_2^\nu 
    -\varepsilon_{\rm L}^\mu\varepsilon_{\rm L}^\nu$.       Thus  $\rho^{\mu\nu}$ is composed of terms $\sim \varepsilon_1^\mu \varepsilon_1^\nu, \varepsilon_2^\mu\varepsilon_2^\nu , \varepsilon_{\rm L}^\mu\varepsilon_{\rm L}^\nu$, as well as $\varepsilon_{\rm 1}^\mu\varepsilon_{\rm L}^\nu$, but not $\varepsilon_{\rm 2}^\mu\varepsilon_{\rm L}^\nu$ by symmetry, or $\tilde q^\mu \tilde q^\nu$ since $q_\mu$ is orthogonal to $\rho^{\mu\nu}$.  With a little algebra we find the general structure,
 \beq
 \rho^{\mu\nu} &=& 
 \varepsilon_{\rm L}^\mu\varepsilon_{\rm L}^\nu\rho^{\rm L} 
 +\varepsilon_1^\mu\varepsilon_1^\nu\rho^{\rm T}_1 
 +\varepsilon_2^\mu\varepsilon_2^\nu \rho^{\rm T}_2 
 + {\cal N}^\mu {\cal N}^\nu  \rho_n  
\nonumber\\
 &= & -(g^{\mu\nu} -\tilde q^\mu \tilde q^\nu)\,\rho^{\rm L} 
  +\varepsilon_1^\mu\varepsilon_1^\nu(\rho^{\rm T}_1-\rho^{\rm L}) 
  +\varepsilon_2^\mu\varepsilon_2^\nu (\rho^{\rm T}_2-\rho^{\rm L}) + {\cal N}^\mu {\cal N}^\nu  \rho_n , 
   \label{impi}
\eeq
where $
  {\cal N}^\mu \equiv n^{\mu} - (\tilde{q}n) \tilde{q}^{\mu}$, with $(ab)\equiv a^\mu b_\mu$.
Compared to the structure of the photon polarization operator in an isotropic medium, the momentum-space anisotropy of the system leads to an extra structure function, $ \rho_n$ term, and in addition a difference of $\rho_1^{\rm T}$ and $\rho_2^{\rm T}$ in general.  

   The new spectral function, $\rho_n$, vanishes when the quark and gluon distribution functions are even under parity, so that $\vec n$ enters only as a special axis, and not as a special direction. Under the successive transformations of parity followed by letting  $\hat n\to -\hat n$, the polarization vector $\vec\varepsilon_{\rm L}$ transforms as a vector, while $\vec \varepsilon_1$ transforms as a pseudovector; thus the cross term $\sim \varepsilon_1\varepsilon_L$ does not occur in $\rho^{\mu\nu}$, and hence neither does $\rho_n$.   In symmetric collisions of two identical nuclei, there should not be a special direction in the local rest frame of the matter; however, for asymmetric collisions, e.g., $A$ on $A'$, 
one expects a non-zero $\rho_n$ term in the photon spectral function.  

  The various scalar spectral functions in Eq.~(\ref{impi}) depend separately on the local $q^0$, $q_\perp$, and $\vec q\cdot \hat n$, where $q_\perp$ is the magnitude of the component of $\vec q$ orthogonal to $\hat n$; or covariantly, they depend  on $Q^2$, 
  $(q u)$, as well as on $(q n)$,  where $u_\mu$ is the 4-velocity of the local rest frame.

      The production rate of dilepton pairs is then proportional to 
  \beq
 \frac{1}{2} \rho^{\mu\nu}L_{\mu\nu} 
  =  Q^2\left(\rho^{\rm T}_1+\rho^{\rm T}_2 + \rho_n\right) + 4m^2\rho^{\rm L}  -s_1^2(\rho^{\rm T}_1-\rho^{\rm L}) - s_2^2(\rho^{\rm T}_2 -\rho^{\rm L}) 
+ \left((qn)^2- (sn)^2\right)\, \rho_n \, , 
\label{saniso}
\eeq 
where we use $Q^2+s^2 = 4 m^2$ and $(sq)=0$, and we define $s_i \equiv (s\varepsilon_i)$ ($i=1,2$), the components of $\vec s$ transverse to $\vec q$ in the local rest frame:  $\vec s_\perp = s_1 \vec\varepsilon_1 + s_2\vec \varepsilon_2$.  This equation gives the dilepton production rate in terms of the projections of $s$ along $\hat \varepsilon_1$ and $\hat\varepsilon_2$, and $n$.
The $s_1^2$ and $s_2^2$ terms  contain the anisotropy produced by transverse virtual photons;  the $s_z ^2$ term arises from the mixing of longitudinal and  transverse ($\vec \varepsilon_1$) virtual photons.    Again, in symmetric collisions with parity invariance in the local matter rest frame, the final $\rho_n$ term is not present.

 Writing $s_1 =|{\vec s}_\perp| \cos \phi_s $ and
$s_2 =  |{\vec s}_\perp| \sin\phi_s$ to bring out the anisotropy,  we see that the production rate becomes
\beq
 \frac{dR_{l^+  l^-}}{d^3\bar{p} d^3\bar{p}' } =  \frac{\alpha^2}{2\pi^4 Q^4}  \left[ 2 Q^2  \bar{\rho}^{\rm T}  + \left(s_\perp^2 + 4m^2  \right)   \rho^{\rm L}  + \left( Q^2 + (qn)^2 - (sn)^2\right)  \rho_n 
 -  |\vec{s}_\perp|^2 \left(   \bar{\rho}^{\rm T} +  \delta{\rho}^{\rm T} \cos 2 \phi_s \right) \right],
\label{saniso2}
\eeq 
where $\bar{\rho}^{\rm T} \equiv ( \rho_1^{\rm T}  + \rho_2^{\rm T} )/2$ and    
$\delta{\rho}^{\rm T} \equiv ( \rho_1^{\rm T}  -  \rho_2^{\rm T} )/2$. The $\cos 2\phi_s$ and $\rho_n$ terms are the effects of anisotropy.
Equation~(\ref{saniso2}) is the principal result of our analysis of the structure of the dilepton rate in the presence of anisotropy.\footnote{Reference~\cite{bengt} arrives at a similiar result, but does not include the anisotropic terms $\delta\rho^T$ and $\rho_n$.}  We emphasize that this structure is valid for virtual photon production processes in all types of collisions, not simply production via interacting quarks and gluons. 

  For $m=0$ with $\rho_n$ absent, we find the dilepton production rate,\footnote{This distribution is similar to that fitted in the NA60 analysis of dimuon pairs produced in In-In collisions at 158 GeV-A at the CERN SPS \cite{NA60}, where there the angles are defined in the Collins-Soper frame;  for NA60 data averaged over all lab directions of the virtual photons,  the corresponding $\lambda$, $\mu$, and $\nu$ are consistent with zero.}
\beq
  \frac{dR_{l^+  l^-}}{d^3\bar{p} d^3\bar{p}' } =\frac{\alpha^2}{2\pi^4 Q^4}\zeta\vec s\,^2\left(1 + \lambda_s \cos^2\theta_s + \mu_s \sin 2 \theta_s \cos\phi_s + (\nu_s/2) \sin^2\theta _s\cos 2 \phi_s\right),
\eeq   
with $ \lambda_s = (\bar{\rho}^{\rm T} -\rho^{\rm L})/\zeta$,   $\nu_s = -2\delta{\rho}^{\rm T}/\zeta$,
and $\mu_s = 0$, where  $\theta_s$ is the angle between
$\vec q$ and $\vec s$, and $\zeta = \bar{\rho}^{\rm T}(1- 2 s_0^2/\vec s\,^2) + \rho^{\rm L}$.

  By comparison the production rate of real photons ($Q^2=0$) of polarization  
 $\varepsilon^\mu$ is \cite{Schenke:2006yp,BH,Bhattacharya:2015ada} is
\beq
\frac{dR_{\gamma}}{d^3 \bar{q}} =
  \frac{\alpha}{2\pi^2} \varepsilon^*_{\mu} \rho^{\mu \nu} \varepsilon_{\nu} 
= \frac{\alpha}{2\pi^2} \left( \bar\rho^{\rm T} + \delta{\rho}^{\rm T} \cos 2 \phi_\varepsilon \right),
\eeq
 where $(\varepsilon\varepsilon_1) \equiv -\cos\phi_\varepsilon$, $(\varepsilon\varepsilon_2) \equiv -\sin\phi_\varepsilon$, and
$d^3\bar{q}= d^3 q / 2| \vec{q}\,|$. 
The anisotropy for real photons arises entirely from the difference, $\delta\rho^{\rm T}$, 
of $\rho^{\rm T}_1$ and $\rho^{\rm T}_2$: the spectral function $\rho_n$ does not enter. 

\begin{wrapfigure}[9]{r}{7cm}
\vspace{-1.2cm}
\hspace{-8pt}
\includegraphics[width=7.0cm]{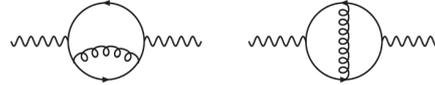}
\vspace{-3cm}
\caption{Structure of the photon polarization processes corresponding to those shown in Fig.~\ref{fig:Compton}.  In a realistic calculation the lines must include hard thermal loops, as in Fig.~\ref{bubbles}. }
\label{nextorder}
\end{wrapfigure} 

\section{Total dilepton rates}

   Equation~(\ref{saniso2}) is not the end of the story.   To derive the total dilepton production rate in an anisotropic ultrarelativistic heavy-ion collision, it is necessary first to calculate the individual spectral functions, $\rho^T_1$. $\rho^T_2$. $\rho^L$, and $\rho_n$ in terms of the space and time quark, antiquark, and gluon distributions present in the collision, and then integrate in space and time over the full collision.   Such a calculation, a generalization of
prior calculations for real photon production in an anisotropic quark-gluon plasma \cite{Schenke:2006yp,Bhattacharya:2015ada}, is currently in progress \cite{future}.  The principle ingredients being included, in addition to simple Drell-Yan processes in the medium \cite{ms-dy,virtual}, are the nominally second order processes in the strong interaction coupling $\alpha_s^2$,  as illustrated in Fig.~\ref{nextorder}, corresponding to Compton and pair annihilation production of virtual photons (cf. Fig.~\ref{fig:Compton}), with full space and time dependent anisotropic quark, antiquark, and gluon distributions, hard thermal loops, and soft scale processes, together with a description of the space-time evolution using full three dimensional anisotropic hydrodynamics \cite{anhydro}.

\section*{Acknowledgments}   This collaboration originated at the workshop at the ECT* in Trento  on ``New perspectives on Photons and Dileptons in Ultrarelativistic Heavy-Ion Collisions at RHIC and LHC" in late 2015.    GB and TH thank the RIKEN iTHES Project and iTHEMS Program for partial support during the course of this work.   This research was also supported in part by NSF Grant PHY1305891, JSPS Grants-in-Aid No.~25287066, and DOE Grant No. DE-SC0013470.  We thank the Aspen Center for Physics, supported in part by NSF Grant PHY1066292, where part of this research was carried out, and thank Y. Akiba, H. Hamagaki, and B. V. Jacak  for 
helpful discussions on the detection of photon polarization in ultrarelativistic heavy-ion collisions.  


\bibliographystyle{elsarticle-num}

\end{document}